# Elementos de ingeniería de explotación de la información: réplica y algunos trazos sobre teoría informática

Rodrigo Lopez-Pablos[1]
rodrigo.lopezpablos@educ.ar
(UNLaM-UTN)

Diciembre de 2013

## 1. El comentario de Gustavo Yana

El comentario de Yana [2013] a un trabajo que propone una introducción de elementos de ingeniería de explotación de información en la administración pública [Lopez-Pablos, 2013], en una contemporaneidad donde se discute el carácter pseudocientífico sobre parte importante de la teoría económica estándar en boga en las últimas generaciones [Bunge 2013], contribuye con su esfuerzo y tiempo a hacer de la nuestra una ciencia más auténtica.

Yana [*op. cit.*] pone de relieve la falta de antecedentes internacionales citados en el comentado trabajo, en sus palabras: «*[. . . ] sería recomendable hacer una revisión de literatura y casos de estudios internacionales,*» proponiendo para ello el análisis de Castellón González y Velásquez [2013], en donde se apoya proponiendo un listado de técnicas. No es difícil rescatarse que la tabla de técnicas metodológicas carece de consistencia taxonómica y explicativa:

[i.] Esta no distingue entre métodos estadísticos triviales –regresiones, series de tiempo– de sistemas inteligentes, cuando expertos del campo advirtieron de la superioridad de los últimos sobre los primeros para el descubrimiento de patrones en base de datos, que como bien afirman los maestros del arte: «*Los métodos basados en sistemas inteligentes, permiten obtener resultados de análisis de la masa de información que los métodos convencionales no logran [. . . ]*» [Britos y Garcia-Martinez, 2009; pág. 1041].

[ii.] La no pormenorización de las técnicas para la detección de fraude para los distintos organismos de administración tributaria no provee de una aproximación algorítmica específica ni respeta una taxonomía pertinente correspondiente a cada uno de los casos, entremezclan técnicas estadísticas con sistemas inteligentes, lo que le quita coherencia a la tabla: *e. g.* en la técnica «redes neuronales» no se distinguen si se trata de algoritmos de una capa (SLP), multicapa (MLP), de Hopfield o Kohonen (SOM); este último, al que no se considera como red neuronal más sino como una técnica independiente contribuyendo a la inconsistencia taxonómica de la tabla, y por consiguiente, confundiendo aún más al experto no informático.

[iii.] Castellón González y Valásquez [*op. cit.*] se enfocan –solamente– en la detección de contribuyentes que falsifican facturación, sin considerar el verdadero potencial de la herramienta computacional; lamentablemente, aquellos no se rescataron de la capacidad de los sistemas inteligentes siendo que su uso apropiado puede detectar patrones organizacionales complejos que se nutren de la ruptura del tejido social que más perjuicio hace a la función benefactora de los Estados.

---

[1] Investigador en ciencias económicas e informáticas. Usuales cláusulas de responsabilidad se aplican.

Por otra parte, en las administraciones tributarias seleccionadas, no se presentan las problemáticas fiscales particulares para cada nación considerada, de este forma, toda combinación de herramental sobreviene en superflua si no se apoya en los casos de aplicación y problemática social de cada comunidad. La situacionalidad única del capital social de cada comunidad está fundamentada en lazos culturales e históricos particulares por lo que difícilmente puedan compararse, al menos sin un profundo análisis descriptivo de cada caso estudiado.

El matiz quizás más discutible del trabajo citado recae en el herramental utilizado, un *SPSS clementine* de código cerrado y licenciado, sobrecapacitado para la labor artesanal, disennado para grandes datawarehouses y haciendo de toda búsqueda de derrame tecnológico-territorial un esfuerzo inútil para el administrador tributario; objetivo este absolutamente fundamental del trabajo comentado, motivo por el cual se descarta dicha cita como antecedente válido.

Desde la interdisciplinariedad informacional entonces, la función tributaria del Estado en finanzas públicas, y la literatura computacional aplicada en la metodología abordada en Lopez-Pablos [*op. cit.*] y Yana [*op. cit.*] no parecen tener conmensurabilidad contemporánea. La metodología fundamentada en Britos y Garcia-Martinez [*op. cit.*] ofrece una técnica para el proceso de explotación de la información *situ* en la frontera del arte computacional actual; ergo, en los trabajos comentados y replicados, se asume un nivel de originalidad epistémico equivalente sobre el campo científico que nos reúne.

Finalmente, la falta de claridad tanto en la distinción de las metodologías como en la aplicación heurística declarada por el comentarista radica probablemente en la falta de formación sistémica e informática la cual nos ha caracterizado a todos los que nos hemos formado como investigadores a partir de una base de teoría económica estándar ya parcial o totalmente obsoleta. En ese orden, se intenta echar un poco de luz en el siguiente apartado.

## 2. La generación de conocimiento desde un punto de vista sistémico

Ilustrando el proceso de información sistémicamente cómo aquel que genera conocimiento a través del tiempo y el actor humano, dado que a partir de la teoría de la información puede verse a cualquier actor o conjunto de estos como generador intrínseco de conocimiento[2], se tiene:

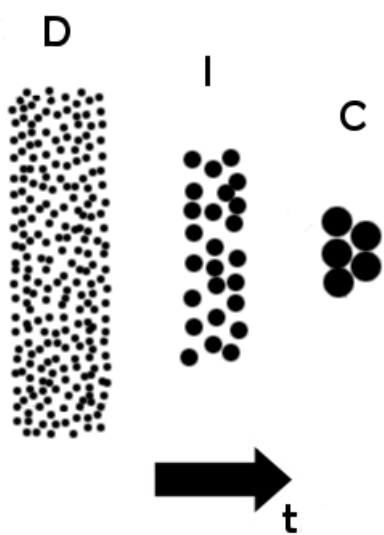

Fig. 1: Generación de conocimiento en el tiempo

Donde se aprecia a los datos «**D**», como nodo de entrada al sistema de información, los que con el paso único e irreversible del tiempo «*t*» conllevan un proceso informativo «**I**» para la generación de un conocimiento «**C**» producto de la interacción heurística última del humano como flujo de salida, proceso para el cual es necesario un tiempo de contemplación mínimo que requiere cierto aislamiento.

---

[2] Tal significación puede adquirir una relevancia civil paradigmática, siendo que a partir de dicha concepción informática, toda persona se constituiría como tal solo a partir del momento en que esta sea capaz de procesar información *per se*. Incluso una nueva taxonomía de la evolución asomaría, pudiéndose catalogar a las especies según la cantidad y calidad de información que produzca cada una; *q. e.*, el código de cómputo entendido cómo una evolución natural del código genético-biológico.

A manera de ejemplo de este proceso, de una cita del comentarista nos cabe que: *«Si el criterio que la elección de atributos a condicionar se sustente a partir de ‹descubrimiento de conocimiento› en cada paso, debería explicitarse, ya que a simple vista no queda claro.»*. De donde puede intuirse un claro ejemplo de falta de contemplación de la información: puesto que no es solo lo que Yana observa, sino el tiempo de contemplación sobre la totalidad del problema lo que hace al aprendizaje, el descubrimiento y la generación del conocimiento.

Siendo que es el tiempo de interacción heurística del humano con los datos en los procesos de información lo que genera el conocimiento; por ende, con bajos niveles de contemplación no habrá aprendizaje ni mucho menos generación de conocimiento. Solo de la observación profunda, con tiempo necesario de contemplación, los instrumentos adecuados, y los datos pertinentes puede haber generación de conocimiento primero para el investigador, y luego para la sociedad a la que este pertenece. Se desprende entonces el siguiente estamento.

**Proposición 1** *La generación de conocimiento se da en la interacción y contemplación heurística del actor humano con la interface informática que este observa –lato sensu–, puesto que por más automatizados que se encuentre, en una última instancia aún hace falta la intervención de un experto humano para darle sentido útil a ese conocimiento.*

Por más automatizados que se encuentre, en una última instancia aún hace falta la intervención de un experto humano para darle sentido útil a ese conocimiento, en una última instancia esta siempre tendrá que ser supervisada, por lo que [aún] no hay una independencia automática completa en su generación como se plasma en la siguiente heurística[3].

$$D \xrightarrow{auto} I \xrightarrow{superv} C$$

Aunque las maquinas gestionan datos y procesan información, aún son incapaces de generar conocimiento útil para el hombre independientemente de la intervención de este; el humano en cambio, necesita sobre todo de un tiempo de contemplación y enfoque de energía observacional sobre un estrato del conocimiento para consecutivamente catapultar así una nueva.

## 2.1. Generación de conocimiento y capital social

Al igual que las problemática de dimensionar el capital social la información también presenta problemas de mensurabilidad, puesto que si bien podemos medir cuantitativamente un volumen de datos de manera certera –en cantidad absoluta de bits– no podemos hacer lo mismo para un cúmulo de conocimiento, el cual se encuentra asociado al humano que lo descubrió u aquél que le dedicó al mismo una parte considerable de tiempo de estudio. Se encuentra ligado por ende a una cualidad nominativa difícil de transferir vinculada solamente hasta ahora al experto humano, por lo que entendemos al mismo como parte de la satisfacción de una necesidad humana trascendente [Lopez-Pablos, 2012] por lo que tenemos el siguiente constructo.

**Proposición 2** *Toda generación de conocimiento útil resulta en la satisfacción de necesidades avanzadas de agentes económicos y sociales fenoménicos lo que resulta en el fortalecimiento del capital social físico y digital de la sociedad.*

Lo que no significa que la distribución de dicho conocimiento no provoque perturbación cuando este es derramado. Por lo que se supone que más allá de cierto umbral, cuando el conocimiento generado por los actores humanos no es asimilado en el campo social al cual el agente fenoménico pertenece se genera quizás algún tipo de perturbación en la sociedad.

---

[3] Una medida de independencia podría ser la cuantificación del tiempo en que las maquinas pudieran seguir procesando información, desde un momento $t = 0$, a partir del cual no haya intervención humana alguna.

La correspondencia nominativa de conocimiento a nivel individual digital parecería distorsionar el entorno social y temporal tanto físico como digital en el que el científico se encuentra inserto, generando perturbación en el entorno del mismo si este último –en una dinámica cíclica entre períodos de aislamiento y presentaciones públicas– no difunde ni aplica los conocimientos descubiertos apropiadamente por el mismo, lo que repercute en el capital social en la forma de energía dilapidada y tiempo perdido en la consolidación de tal tejido[4]. Construcción de donde se condensa la siguiente proposición.

**Proposición 3** *El incremento de conocimiento más allá de algún umbral indefinido, así como la acumulación del mismo nominativamente en un agente en particular; al ser este socializado, parece generar una curvatura en el tiempo y espacios digitales y sociales creando perturbación en el entorno del agente fenoménico.*

Del cual podría desprenderse una discusión adicional a la existente en las academias en sí las obras plasmadas en revistas, journals, software, etc. cerradas en acceso, puedan ser considerados o no como conocimiento según su grado de disponibilidad, dado que no socializan la producción científica o al menos retardan la misma al obstaculizar su visibilidad adicionando una restricción adicional al investigador[5]. En una posterior contabilidad del valor y del tiempo podría plasmarse como una pérdida de bienestar para la sociedad toda, puesto que se desaprovecha la satisfacción de necesidades trascendentes por parte de los agentes fenoménicos locales intervinientes en la formación de capital social de un espacio, dando lugar en consecuencia, a capas trascendentes de otras territorialidades ajenas a la originaria; de igual utilidad, pero de menor satisfacción fenoménica.

# Referencias

---

[4] Es importante mencionar que antes del computador y la internet, las publicaciones tradicionales eran los únicos medios que posibilitaban el derrame social de la ciencia y la tecnología, los cuales se trasmitían a elites con grandes capacidades memorísticas y económicas, únicas capaces en aquel tiempo -al contar con la energía y el tiempo necesario- de derramar y trasmitir conocimiento a través de las generaciones.

[5] Restricción adicional al que *de facto* representa el tiempo de contemplación necesario para producir conocimiento, así como la posibilidad física de acceso a internet, barreras estructurales a ser suplidas al menos parcialmente por los Estados.